\documentclass{article}
\usepackage{spconf,amsmath,amsfonts,graphicx,cite, mathtools, comment}

\newcommand{\sub}[2]{#1_\mathrm{#2}}

\newcommand{\ex}[1]{\mathbb{E}\left[ #1 \right]}
\newcommand{\vb}[1]{\mathbf{#1}}
\newcommand{\vsub}[2]{\mathbf{#1}_\mathrm{#2}}
\newcommand{\rmsub}[2]{\mathrm{#1}_\mathrm{#2}}

\DeclarePairedDelimiter\floor{\lfloor}{\rfloor}

\title{ACOUSTIC INTENSITY, ENERGY-DENSITY AND DIFFUSENESS ESTIMATION IN A DIRECTIONALLY-CONSTRAINED REGION}
\twoauthors
  {Archontis Politis}
	{Aalto University\\
	Department of Signal Processing and Acoustics\\
	Espoo, Finland}
  {Ville Pulkki}
	{Aalto University\\
	Department of Signal Processing and Acoustics\\
	Espoo, Finland}
\begin{document}

\maketitle
\begin{abstract}
This work presents a method for estimation of the acoustic intensity, the energy density and the associated sound field diffuseness around the origin, when the sound field is weighted with a spatial filter. The method permits energetic DOA estimation and sound field characterization focused in a specific angular region determined by the beam pattern of the spatial filter. The formulation of the estimators is presented and the their behavior is analyzed for the fundamental cases useful in parametric sound field models of a single plane wave, a uniform diffuse field and a mixture of the two.
\end{abstract}
\section{Introduction}
\label{sec:intro}

Acoustic intensity is a physical quantity with fundamental applications in acoustical engineering. It expresses the acoustical energy flow for a point through a unit surface, and as such it s time and space distribution is a powerful indicator of the acoustic field properties and the sources that generate it. It's time-averaged version, termed active acoustic intensity expresses the mean flow of energy and hence it is an even more useful indicator of strong directional components in the field since it ignores the effect of components that produce no mean flow of energy such as reactive components close to sources or standing waves.

From the wide variety of applications of acoustic intensity (for example measurement of absorption coefficients, source power etc.), this work relates mostly to the ones related to identification and localization of sources in the field.  For a point source or plane wave in the presence of diffuse noise, the active intensity vector has the opposite direction of the direction-of-arrival (DOA) of the source. This capability has been studied and exploited for a variety of applications such as localization and tracking \cite{Jarrett2010, thiergart2009localization, Dimoulas2009}, source separation \cite{Gunel2008, Bunting2013} and spatial audio coding \cite{pulkki2007spatial, Hurtado-Huyssen2005}. Furthermore, its statistical distribution can reveal the presence and DOAs of multiple sources as has been shown by \cite{Tervo2009, Gunel2013}.

Another interesting property of the acoustic intensity and the energy density is their usefulness for sound field characterization \cite{Jacobsen1990, Gauthier2014}. The ratio of the active intensity to the time-averaged energy-density can reveal if the field is close to an ideal diffuse field or close to a plane-wave field \cite{Jacobsen1989, Schiffrer1994, Vigran1988}. This property has been also exploited in spatial audio coding for assessing the diffuseness of a recorded sound scene or impulse response \cite{DelGaldo2012, Karjalainen2001, pulkki2007spatial}.

Intensity is usually determined by measuring the pressure at the origin and the velocity through the pressure gradient across the three-cartesian axes, by employing some microphone array. A minimum of four microphones on a tetrahedron is required for a three-dimensional measurement of the vector. It can be also measured by the B-format set of signals, which are formed by an omni and three dipole sensors oriented orthogonally. Measurement of the velocity components directly is also possible through anemometric probes such as the Microflown \cite{Cengarle2011}. A comprehensive analysis of measurement of the intensity vector with microphone arrays is presented in \cite{Hacihabiboglu2013}.

In this work we consider a formulation that allows to measure the intensity and energy density of a weighted sound field by some beam-pattern. That allows an energetic sound field  analysis of angular regions, since sources that fall outside the main lobe of the beam pattern are attenuated significantly and their effect on the intensity vector, energy density and diffuseness can be insignificant. Hence, applications mentioned above can be performed with local regional descriptors instead of global ones. The formulation is presented in the spherical harmonic domain which provides a convenient framework for obtaining closed-form solutions to the spatially weighted energetic quantities, with respect to the coefficients that define the spatial filter. Furthermore, the modified quantities are analyzed for the basic field cases of a plane wave, a diffuse field and a mixture of the two. The proposed directionally-constrained energetic analysis has been applied by the authors to high-resolution parametric analysis and synthesis of spatial sound scenes \cite{pulkki2013parametric, politis2015sector}. Further potential applications may include analysis and synthesis of spatial room impulse responses, perceptually-motivated spatial audio coding of sound fields, and directional sharpening of sound distributions for improved acoustic analysis and DoA estimation.

\section{Background}
\label{sec:background}

Let us consider an incident plane wave field captured or estimated at some radius $R$ around the recording point, expressed by the complex amplitude density of $a(f, \Omega)$, where $\Omega = (\theta, \varphi)$ with inclination from the north pole $\theta\in[0, \pi]$ and azimuth $\varphi \in[-\pi/2, \pi/2]$. The pressure and the acoustic velocity at the origin due to the amplitude distribution is given by
\begin{eqnarray}
	p(f) &=& \int_{\Omega\in S^2} \!a(f, \Omega) \; \mathrm{d}\Omega \label{eq:pressure}\\
	\vb{u}(f) &=& -\frac{1}{Z_0} \int_{\Omega} \! a(f, \Omega) \vb{n}(\Omega) \; \mathrm{d}\Omega = -\frac{1}{Z_0} \vb{v}(f) \label{eq:velocity}
\end{eqnarray}
with integration on the unit sphere $S^2$ denoted as $\int_{\Omega \in S^2} \mathrm{d}\Omega = \int_{-\pi}^{\pi}\int_{0}^{\pi}  \sin\theta \,\mathrm{d}\theta \,\mathrm{d}\varphi$ and $\vb{n}(\Omega) = [\sin\theta \cos\varphi,\, \sin\theta \sin\varphi,\, \cos\theta]^T$ the unit vector pointing to the direction of incidence. The signal vector $\vb{v}(f) =   \left[\rmsub{v}{x}(f),\rmsub{v}{y}(f),\rmsub{v}{z}(f)\right]^T$ corresponds to minus the unnormalized cartesian components of the particle velocity and $Z_0=c \rho_0$ is the characteristic impedance of air.

The spherical harmonic transform (SHT) of the amplitude function is given by 
\begin{equation}
	a_{nm}(f) = \int_{\Omega} \! a(f, \Omega) Y^*_{nm}(\Omega) \;\mathrm{d}\Omega
\end{equation}
where $Y_{nm}$ are the spherical harmonics (SHs) of order $n$ and degree $m$, and $(^{*})$ denotes complex conjugation. The spherical harmonics are defined as
\begin{equation}
	\label{eq:shs}
	Y_{nm}(\Omega) = \sqrt{\frac{2n+1}{4\pi}\frac{(n-m)!}{(n+m)!}} P_n^{m}(\cos\theta) e^{im\varphi}
\end{equation}
where $P_n^{m}$ are associated Legendre functions of degree $n$ and order $m$. We can compress the two indices $(n,m)$ of the SHs to a single one $q$, with an one-to-one relation between them, where $q=n(n+1)+m$ and inversely $n = \lfloor \sqrt{q} \rfloor$ and $m=n^2-q$. This notation will be used in the present work for brevity, apart from cases where it is clearer to show explicitly the degree and order of the spherical functions. 

For an angularly band-limited function $f(\Omega)$ up to some order $N$, its vector of spherical harmonic coefficients is denoted as $\vb{f}_N = [f_0,\, f_1,\, ..., f_{Q-1}]^T$, with $Q = (N+1)^2$. The following properties of the SHT will be used extensively in this work. First, the SHs of (\ref{eq:shs}) are orthonormal, meaning that
\begin{equation}
	\int_{\Omega} \! Y_{q'}(\Omega) Y^*_{q}(\Omega) \; \mathrm{d}\Omega = \delta_{q-q'}
\end{equation}
where $\delta_{q-q'}$ is the Kronecker delta function. Second, based on this orthonormality relation it is easy to show that for two band-limited spherical functions $f(\Omega)$ and $g(\Omega)$ the following relations hold
\begin{eqnarray}
	\int_{\Omega} \! f(\Omega) g^*(\Omega) \;\mathrm{d}\Omega = \displaystyle\sum_{q=0}^{Q-1} f_q\, g_q^* =  \vb{g}^H_N \, \vb{f}_N \label{eq:parceval2}\\
	\int_{\Omega} \! f(\Omega) f^*(\Omega) \;\mathrm{d}\Omega = \displaystyle\sum_{q=0}^{Q-1} |f_q|^2  = \vb{f}^H_N \, \vb{f}_N \label{eq:parceval},
\end{eqnarray}
which are cases of Parseval's theorem for the SHT.

By applying a spatial filter of order $N$ on the incident plane waves defined by the real directional pattern $w(\Omega)$, we can estimate the output in the SH domain as in (\ref{eq:parceval2})
\begin{align}
	\label{eq:beam}
	\sub{p}{w}(f) = \int_{\Omega} \! w(f, \Omega) a(f, \Omega) \; \mathrm{d}\Omega = \vb{w}_N^H(f) \, \vb{a}_N(f)
\end{align}
where $\vb{w}_N$ is the vector of the $(N+1)^2$ SH coefficients of $w(\Omega)$. A special case of interest in this work are real rotationally symmetric beam-patterns $c(\Omega) = c(\theta)$ oriented at the $z$-axis, which are fully described by $N+1$ real SH coefficients $\vb{c}_N = [c_0,\, c_1,\, ..., c_N]^T$ of degree $m=0$. Such a pattern can be easily steered to an arbitrary direction $\Omega_0$, in which case all its $(N+1)^2$ SH coefficients are populated as in
\begin{align}
	\label{eq:steer}
	w_{nm}(f) = \sqrt{\frac{4\pi}{2n+1}}c_n \, Y_{nm}^*(\Omega_0).
\end{align}

\section{Directionally weighted Intensity and Diffuseness Estimation }

In order to estimate the intensity and energy density, the pressure and particle velocity at the origin are needed. The pressure can be captured with an omnidirectional pattern, while the velocity, as it is obvious from (\ref{eq:velocity}), can be captured with three dipole patterns $x(\Omega), \, y(\Omega), \, z(\Omega)$ corresponding to the components of $\vb{n}(\Omega)$ as in
\begin{equation}
\label{eq:dipole}
\vb{n}(\Omega) =  \left[ {\begin{array}{c}
 	x(\Omega) \\
	y\Omega) \\
	z(\Omega)
	\end{array} } \right] = \
\left[ {\begin{array}{c}
 	\sin\theta \cos\varphi\\
	\sin\theta \sin\varphi\\
	\cos\theta
	\end{array} } \right]
\end{equation}
The signals captured by these orthogonal dipoles form the signal vector $\vb{v}(f)$, and together with the omnidirectional pressure signal $p(f)$ are known in spatial sound recording and reproduction literature as the B-format signal set.

From the pressure and velocity signals, the active intensity vector for a monochromatic field $\vsub{i}{a}(f)$ can be measured as \cite{Fahy1995}
\begin{align}
\label{eq:intense}
\vsub{i}{a}(f) = \frac{1}{2} \Re \left\{ p^*(f) \, {\vb{u}(f)} \right\} = -\frac{1}{2Z_0} \Re \left\{ p^*(f) \, {\vb{v}(f)} \right\}.
\end{align}
and the energy density of the sound field at the same point as \cite{Fahy1995}
\begin{align}
\label{eq:edensity}
E(f) &= \frac{\rho_0}{4} \| \vb{u}(f) \| ^2+\frac{1}{4\rho_0c^2}|p(f)|^2 \notag\\
		&= \frac{1}{4\rho_0c^2}\left[||\vb{v}(f)||^2+|p(f)|^2\right].
\end{align}
Finally, the diffuseness is defined as \cite{DelGaldo2012}
\begin{equation}
\label{eq:diffuseness}
	\psi(f) = 1 - \frac{\|  \vb{i}_a(f) \|}{c\, E(f)} = 1-\frac{2||  \Re\{p^*(f)\, \vb{v}(f)\}||}{ ||\vb{v}(f)||^2+|p(f)|^2 }.
\end{equation}

By applying a spatial filter described by the pattern $w(\Omega)$, we impose a directional weighting on the amplitude density function. The weighted pressure signal is given by (\ref{eq:beam}) while the velocity signals due to the same weighted distribution are
\begin{equation}
	\label{eq:beam_vel}
	\vsub{u}{w}(f) = -\frac{1}{Z_0}\int_{\Omega} \! \mathrm{w}(f, \Omega)\vb{n}(\Omega)\, a(f, \Omega)  \;\mathrm{d}\Omega = -\frac{1}{Z_0}\vsub{v}{w}(f).
\end{equation}
The signal vector $\vsub{v}{w}$ corresponds to the signals captured with the beam patterns 
\begin{equation}
\label{eq:beam_xyz}
w(\Omega) \vb{n}(\Omega) = \left[ {\begin{array}{c}
 	\sub{w}{x}(\Omega) \\
	\sub{w}{y}(\Omega) \\
	\sub{w}{z}(\Omega)
	\end{array} } \right] = 
\left[ {\begin{array}{c}
 	w(\theta, \varphi)\sin\theta \cos\varphi \\
	w(\theta, \varphi) \sin\theta \sin\varphi \\
	w(\theta, \varphi) \cos\theta
	\end{array} } \right].
\end{equation}
It is evident from (\ref{eq:beam_vel}) that it in order to capture the velocity components $\vsub{u}{w}$ of the weighted field, beam patterns that are products of the original pattern and the three orthogonal dipoles as in (\ref{eq:beam_xyz}) should be generated. These velocity patterns obey the following energy preserving property.
\begin{equation}
	\label{eq:velproperty}
	\sub{w}{x}^2(\Omega)+\sub{w}{y}^2(\Omega)+\sub{w}{z}^2(\Omega)=w^2(\Omega).
\end{equation}

Based on the weighted pressure and velocity signals, it is possible to measure the three energetic quantities of interest in a non-global manner but with directional selectivity. More specifically, a local active intensity $\vsub{i}{w}$, energy density $\sub{E}{w}$ and local diffuseness $\sub{\psi}{w}$ can be estimated with the exact same formulas of (\ref{eq:intense})-(\ref{eq:diffuseness}), by replacing $p,\, \vb{v}$ with their weighted versions $\sub{p}{w},\, \vsub{v}{w}$. Considering symmetric sector beams with a main lobe, sounds incident from directions close to the sector orientation contribute mostly to the estimated parameters, while sounds far from it, contribute from little to negligible, depending on the sector beam pattern and width.

\section{Higher-order input and pattern generation}
The problem of the estimation of the energetic quantities in a spatially weighted field becomes a problem of generating the patterns of (\ref{eq:beam_xyz}). This task can be conveniently formulated in the spherical harmonic domain (SHD) in which a closed form solution for the beam-forming weights resulting in these patterns can be derived. The shape and orientation of the spatial filter $w(\Omega)$ is a design choice which translates to the pattern's coefficient vector $\vb{w}_N$ in the SHD, where $N$ is the SH order of the beam-pattern assuming that is directionally band-limited. Its corresponding velocity patterns then are of order $N+1$ since they are products of the original pattern and the dipole components of $\vb{n}(\Omega)$ which are of order $N=1$. If the measurement system can support orders up to $N+1$, then the field amplitude SH coefficients $a_{N+1}$ can be captured and the weighted pressure and velocity can be measured from (\ref{eq:beam}) and (\ref{eq:beam_vel}) by
\begin{align}
\left[ \begin{array}{c} 
	p_\mathrm{w}\\
	\vb{v}_\mathrm{w} \end{array} \right] = [\vb{w}_{N+1}, \vb{w}^\mathrm{x}_{N+1}, \vb{w}^\mathrm{y}_{N+1}, \vb{w}^\mathrm{z}_{N+1}]^T \vb{a}_{N+1}
\end{align}
where $\vb{w}_{N+1}^\mathrm{x},\, \vb{w}_{N+1}^\mathrm{y},\, \vb{w}_{N+1}^\mathrm{z}$ are the coefficients of the velocity patterns. These coefficients can be expressed as linear combinations of the pattern coefficients $\vb{w}_N$, determined by $(N+2)^2\mathrm{x}(N+1)^2$ matrices $\vb{A}_\mathrm{x},\ \vb{A}_\mathrm{y},\ \vb{A}_\mathrm{z}$ as in
\begin{equation}
	\label{eq:velbeamSH}
	\vb{w}^\mathrm{x,y,z}_{N+1} = \vb{A}_\mathrm{x,y,z} \, \vb{w}_N.
\end{equation}
The members of the matrices $\vb{A}_\mathrm{x,y,z}$ are deterministic and independent of the spatial filter's pattern, they depend on the $1st$-order SH coefficients of the $\vb{n}(\Omega)$ and the order $N$. They can be found analytically from a combination of Gaunt coefficients and the coefficients of the three orthogonal dipoles expanded in complex SH. The three dipoles $x,\, y,\, z$ of $\vb{n}(\Omega)$ expanded on the complex SH series result in the following coefficients of order $n=1$ or equivalently $q=1-3$
\begin{align}
\label{eq:dipoleSH}
[\vb{x}_1,\, \vb{y}_1,\, \vb{z}_1] = \left[ \begin{array}{ccc} 
							\sqrt{\frac{2\pi}{3}} 	&	\sqrt{\frac{2\pi}{3}}i	&	0\\
							0				&	0				&	\sqrt{\frac{4\pi}{3}}\\
							-\sqrt{\frac{2\pi}{3}}	&	\sqrt{\frac{2\pi}{3}}i	&	0
							\end{array} \right]
\end{align}
The members of the matrices for the velocity coefficients are then given by
\begin{eqnarray}
	\label{eq:velmat}
	\floor{\vb{A}_\mathrm{x}}_{ij} &=& x_1 G_{(j-1)1}^{(i-1)} + x_3 G_{(j-1)3}^{(i-1)} \nonumber\\
	\floor{\vb{A}_\mathrm{y}}_{ij} &=& y_1 G_{(j-1)1}^{(i-1)} + y_3 G_{(j-1)3}^{(i-1)} \nonumber\\
	\floor{\vb{A}_\mathrm{z}}_{ij} &=& z_2 G_{(j-1)2}^{(i-1)}
\end{eqnarray}
where $G_{q'q''}^q$ is the Gaunt coefficient relating the $q'$ and $q''$ coefficient of two spherical functions under multiplication, to the $q$ coefficient of their product. A derivation of (\ref{eq:velmat}) is given in the appendix. The matrices of (\ref{eq:velmat}) can be precomputed analytically for up to some desired analysis order $N+1$. Then the velocity patterns for any spatial filter up to order $N$ can be found directly by (\ref{eq:velbeamSH}).

\section{Reference field conditions}

For sound field characterization and DOA estimation, the statistics of the energetic quantities mentioned above are used \cite{DelGaldo2012}. In a single plane wave field the mean active intensity points to the propagation of the wave while in a diffuse field it vanishes. Another widely used quantifier in this context is the diffuseness coefficient which is bounded between $\psi \in [0, 1]$, for a plane wave and a purely diffuse field respectively. In the following, we present the case of a local weighted analysis for these reference field conditions and how they relate to the global energetic quantities.

The following statistical sound field model is used for the derivation. The plane wave component of the sound field is incident from direction $\Omega_l$ carrying a signal $\sub{a}{pw}(f)$ with a power spectral density (PSD) of $\sub{P}{pw}(f) = \ex{|\sub{a}{pw}(f)|^2}$. An isotropic diffuse field is modeled as an amplitude density function $\sub{a}{df}(f,\Omega)$ with the following property
\begin{eqnarray}
\label{eq:diffield}
	\ex{\sub{a}{df}^*(f, \Omega_1)\, \sub{a}{df}(f, \Omega_2)} = \left\{\begin{array}{c}
		\frac{\sub{P}{df}(f)}{4\pi},\; \text{for } \Omega_1=\Omega_2\\
		0,\; \text{otherwise}
	\end{array}\right .
\end{eqnarray}
producing a diffuse pressure signal $\sub{s}{df}(f) = \int_{\Omega} \! \sub{a}{df}(f, \Omega) \, \mathrm{d}\Omega$. Based on the property of (\ref{eq:diffield}), the PSD of the diffuse field is $\ex{|\sub{s}{df}(f)|^2} = \sub{P}{df}(f)$. Furthermore, the plane wave signal and the diffuse sound signal are uncorrelated $\ex{\sub{a}{pw}^*\, \sub{s}{df}} = \ex{\sub{a}{pw}^*\, \sub{a}{df}} = 0$.

The energetic quantities, based on the general relations of (\ref{eq:intense})-(\ref{eq:diffuseness}) can be reformulated for the statistical field description as
\begin{align}
\label{eq:intense2}
\vsub{i}{a}(f) &= -\frac{1}{2Z_0} \Re \left\{\vsub{s}{pv}(f)\right\}\\
\label{eq:edensity2}
E(f) &= \frac{1}{4\rho_0c^2}\left[\sub{S}{vv}(f)+\sub{S}{pp}(f)\right]\\
\psi(f) &= 1-\frac{2 || \Re \left\{\vsub{s}{pv}(f)\right\} ||}{\sub{S}{vv}(f)+\sub{S}{pp}(f)}
\end{align}
where $\sub{S}{pp} = \ex{|p(f)|^2}$ the PSD of the pressure signal and $\sub{S}{vv} = \ex{\vb{v}^H(f) \vb{v}(f)}$ the combined PSDs of the velocity signals. The cross spectral density (CSD) between pressure and velocity is denoted as the vector $\vsub{s}{pv} = \ex{p^*(f) \vb{v}(f)}$.

\subsection{Single plane wave}

For a single plane wave incident from angle $\Omega_l$ the intensity and energy density are 
\begin{eqnarray}
	\label{eq:intensepw}
	\vsub{i}{pw} &=& -\frac{1}{2Z_0} \sub{P}{pw} \vb{n}(\Omega_l)\\
	\sub{E}{pw} &=& \frac{1}{2\rho_0c^2} \sub{P}{pw}.
\end{eqnarray}

If the spatial filter $w(\Omega)$ is applied, the pressure and velocity PSDs, using the property of the velocity patterns (\ref{eq:velproperty}), are simply $\sub{S}{pp,w}(f) = \sub{S}{vv,w} = w^2(\Omega_l)\sub{P}{pw}$ and the CSD is $\vsub{s}{pv,w} = w^2(\Omega_l)\sub{P}{pw} \vb{n}(\Omega_l)$. The active intensity vector and the energy density are then
\begin{eqnarray}
	\label{eq:intensepww}
	\vsub{i}{pw,w} &=& -\frac{1}{2Z_0} w^2(\Omega_l) \sub{P}{pw} \vb{n}(\Omega_l) =  w^2(\Omega_l) \vsub{i}{pw} \\
	\label{eq:edensitypww}	
	\sub{E}{pw,w} &=&  \frac{w^2(\Omega_l)}{2\rho_0c^2} \sub{P}{pw} = w^2(\Omega_l) \sub{E}{pw}.
\end{eqnarray}
From (\ref{eq:intensepww}) it is evident that the modified intensity vector is pointing to the direction of propagation as in the non-weighted case. Furthermore, the diffuseness coefficient is not affected by the presence of the beam and remains zero $\sub{\psi}{pw,w} = 0$.

\subsection{Uniform diffuse field}

For a uniform diffuse field, the CSD between pressure and velocity approaches zero $\vsub{s}{pv}=0$, while the PSDs of pressure and sum of velocity signals are equal to the diffuse field power $\sub{S}{pp} = \sub{S}{vv} = \sub{P}{df}$. Based on these observations, the global energetic quantities for a diffuse field result in the following relations
\begin{eqnarray}
	\vsub{i}{df} &=& 0\\
	\sub{E}{df} &=&  \frac{1}{2\rho_0c^2} \sub{P}{df}\\
	\sub{\psi}{df} &=& 1.
\end{eqnarray}

If a directional weighting is applied on the diffuse field, the mean intensity does not vanish and the upper bound of diffuseness is lower than one. Below the expected values of intensity, energy density and diffuseness are derived with respect to the spatial filter's coefficients $\vb{w}_{N}$ of order $N$. For the derivation, we define the velocity pattern matrix $\vb{W}_\vb{n} = [\vb{w}^{\mathrm{x}}_{N+1}, \vb{w}^{\mathrm{y}}_{N+1}, \vb{w}^{\mathrm{z}}_{N+1}]$. A useful quantity is the mean of the DOA vector $\vb{n}(\Omega)$ averaged over the sphere and weighted with the function $w^2(\Omega)$ which based on (\ref{eq:parceval2}) and (\ref{eq:velbeamSH}) is 
\begin{eqnarray}
\label{eq:modn}
	\vb{k} = \int_{\Omega} \! w^2(\Omega)\vb{n}(\Omega) \;\mathrm{d}\Omega = \vb{w}^H_{N+1} \vb{W}_\vb{n}
\end{eqnarray}
where $\vb{w}_{N+1}$ is the spatial filter's coefficient vector zero-padded to length $(N+2)^2$. Another useful quantity is the spatial filter's directivity factor which based on (\ref{eq:parceval}) is
\begin{eqnarray}
\label{eq:dirfactor}
	Q(f) = \frac{4\pi}{\int_{\Omega} \! w^2(\Omega) \;\mathrm{d}\Omega} = \frac{4\pi}{\vb{w}^H_{N}\vb{w}_{N}}.
\end{eqnarray}

Based on (\ref{eq:parceval}), (\ref{eq:diffield}) and the energy preserving property of the velocity patterns (\ref{eq:velproperty}), the PSDs of the pressure and the combined velocity signals are equal as in
\begin{align}
	\label{eq:pp,w}
	\sub{S}{pp,w} &= \frac{\sub{P}{df}}{4\pi} \int_{\Omega} \!  w^2(\Omega)  \;\mathrm{d}\Omega \nonumber\\
	 \sub{S}{vv,w} &= \frac{\sub{P}{df}}{4\pi} \int_{\Omega} \!  \sub{w}{x}^2(\Omega)+\sub{w}{y}^2(\Omega)+\sub{w}{z}^2(\Omega)  \;\mathrm{d}\Omega \nonumber\\
	 &= \frac{\sub{P}{df}}{4\pi} \vb{w}_N^H \vb{w}_N = \frac{1}{Q}\sub{P}{df}.
\end{align}
The CSD between pressure and velocity, based on (\ref{eq:parceval}), (\ref{eq:diffield}) and (\ref{eq:modn}) is
\begin{align}
	\label{eq:pv,w}
	\sub{\vb{s}}{pv,w} = \frac{\sub{P}{df}}{4\pi} \int_{\Omega} \!  w^2(\Omega)\vb{n}(\Omega)  \;\mathrm{d}\Omega = \frac{\sub{P}{df}}{4\pi} \vb{k}.
\end{align}
The expected intensity and energy density can now be found from (\ref{eq:intense2}), (\ref{eq:edensity2}), (\ref{eq:dirfactor})-(\ref{eq:pv,w}) as
\begin{eqnarray}
	\label{eq:intensediffw}
	\sub{\vb{i}}{df,w} &=& -\frac{\sub{P}{df}}{8\pi Z_0} \vb{k}\\
	\sub{E}{df,w} &=& \frac{1}{2\rho_0c^2 Q} \sub{P}{df} = \frac{1}{Q} \sub{E}{df}.
\end{eqnarray}
Finally, the local diffuseness in a diffuse field, using (\ref{eq:modn})-(\ref{eq:pv,w}), is given by
\begin{equation}
	\label{eq:diffdiffw}
	\sub{\psi}{df,w} = 1-\frac{||\vb{w}^H_{N+1} \vb{W}_\vb{n}||}{\vb{w}_N^T \, \vb{w}_N} = 1 - \frac{Q}{4\pi} ||\vb{k}||.
\end{equation}

In the case of an axisymmetric spatial filter, which would normally be of more practical interest, described exactly by its $(N+1)$-length coefficient vector $\vb{c}_N$ and oriented towards $\Omega_0$, the following observations can be made. First, since the diffuse field is uniform, the orientation of the beam pattern does not matter for the statistics of the energetic quantities and they can be express with respect to the unrotated pattern $c(\theta)$ or its respective coefficients $\vb{c}_N$, without loss of generality. Second, due to the symmetry of the pattern, the weighted average vector of ($\ref{eq:modn}$) points to the direction of the beam pattern $\vb{k} = K \vb{n}(\Omega_0)$, where $K = ||\vb{k}||$. Thus the diffuse field intensity of ($\ref{eq:intensediffw}$) points to the opposite of the beam direction. The magnitude of the $\vb{k}$ vector can be found by
\begin{eqnarray}
\label{eq:modn2}
	K = \vb{c}_{N+1}^T \vb{c}_{N+1}^\mathrm{z}
\end{eqnarray}
where $\vb{c}_{N+1}^\mathrm{z}$ are the $(N+2)$ coefficients of the also axisymmetric $c_\mathrm{z}(\theta) =c(\theta) z(\theta)= $ velocity pattern and can be found by (\ref{eq:velbeamSH}). The reason that  $c(\theta) x(\Omega), c(\theta) y(\Omega)$ do not contribute to the magnitude of $\vb{k}$ is that they vanish after integration in (\ref{eq:modn}).

\subsection{Mixture of plane wave and diffuse field}
In the case of a mixture between a uniform diffuse field and a plane wave, diffuseness is determined by the power ratio between the plane wave and the diffuse field, termed direct-to-diffuse ratio (DDR) $\Gamma = \sub{P}{pw}/\sub{P}{df}$. Since the directional and the diffuse signals are uncorrelated, the mean intensity vector and energy density are the sum of the individual contributions due to the plane wave and the diffuse field of the previous cases
\begin{eqnarray}
	\label{eq:intensepd}
	\vsub{i}{pd} &=& \vsub{i}{pw} = -\frac{1}{2Z_0} \sub{P}{pw} \vb{n}(\Omega_l)\\
	\sub{E}{pd} &=& \sub{E}{pw} + \sub{E}{df} = \frac{1}{2\rho_0c^2} \left[ \sub{P}{pw} + \sub{P}{df} \right] \\
	\sub{\psi}{pd} &=& 1-\frac{\sub{P}{pw}}{\sub{P}{pw} + \sub{P}{df}} = \frac{1}{1+\Gamma}
\end{eqnarray}
where the last relation is derived by division of the fraction by $\sub{P}{df}$. The fact that the intensity of (\ref{eq:intensepd}) in the mixed-field case is the same as in the plane wave case of (\ref{eq:intensepw}) makes it a robust single source DOA estimator in reverberant conditions that resemble an ideal diffuse field.

Under the presence of a spatial filter diffuseness naturally becomes dependent, apart from the DDR, also on the directivity factor of the beam pattern and the DOA of the plane wave. The PSDs and CSD of pressure and velocity signals are simply the sum of the contributions of the plane wave and diffuse field as in
\begin{align}
	\label{eq:pdpp,w}
	\sub{S}{pp,w} &= \sub{S}{vv,w} = w^2(\Omega_l)\sub{P}{pw} + \frac{\sub{P}{df}}{Q} \\
	\label{eq:pdpv,w}
	\sub{\vb{s}}{pv,w} &= w^2(\Omega_l)\sub{P}{pw} \vb{n}(\Omega_l) + \frac{\sub{P}{df}}{4\pi} \vb{k}
\end{align}
and the same is true for the intensity and energy density
\begin{align}
	\label{eq:intensepdw}
	\vsub{i}{pd,w} &= -\frac{1}{2Z_0} \left[  w^2(\Omega_l)\sub{P}{pw} \vb{n}(\Omega_l) + \frac{\sub{P}{df}}{4\pi} \vb{k} \right]\\
	\label{eq:energypdw}
	\sub{E}{pd,w} &= \frac{1}{2\rho_0c^2}\left[ w^2(\Omega_l)\sub{P}{pw} + \frac{\sub{P}{df}}{Q} \right].
\end{align}
Finally, the local diffuseness in the mixed field case becomes
\begin{align}
	\label{eq:diffpdw}
	\sub{\psi}{pd,w} &= 1-\frac{||w^2(\Omega_l)\sub{P}{pw} \vb{n}(\Omega_l) + \sub{P}{df}\vb{k}/(4\pi) ||}{w^2(\Omega_l)\sub{P}{pw} + \sub{P}{df}/Q} \nonumber\\
	&= 1-\frac{||\Gamma w^2(\Omega_l) \vb{n}(\Omega_l) + \vb{k}/(4\pi)||}{\Gamma w^2(\Omega_l)  + 1/Q}.
\end{align}

If the beam pattern is axisymmetric and $\vb{k} = K \vb{n}(\Omega_0)$, it is convenient to parametrize the diffuseness as a function of the DDR and the angle $\alpha$ between the sector orientation $\Omega_0$ and the DOA of the plane wave $\Omega_l$, which is given by the relation $\cos\alpha = \cos\theta_l\,\cos\theta_0 + \sin\theta_l\,\sin\theta_0\,\cos(\varphi_l-\varphi_0)$. After using the law of cosines in (\ref{eq:diffpdw}) we get
\begin{align}
	\sub{\psi}{pd,w}(\Gamma, \alpha) = 1-\frac{\Delta(\Gamma, \alpha)}{\Gamma c^2(\alpha) + 1/Q}
\end{align}
with
\begin{align}
	\label{eq:deltapdw}
	\Delta(\Gamma, \alpha) = \sqrt{\Gamma^2 c^4(\alpha) + (K/4\pi)^2 + 2\Gamma (K/4\pi) c^2(\alpha) \cos\alpha}.
\end{align}
The following observations can be made. In the case that the plane wave is absent $\Gamma=0$ or that its direction coincides with a null of the spatial filter's pattern $w(\Omega_l)=c(\alpha)=0$, the local diffuseness becomes maximum and equal to the diffuse field one of (\ref{eq:diffdiffw}). Furthermore, assuming that the maximum of the spatial filter is normalized to unity, if the direction of the plane wave is the same as the beam orientation $c(\alpha) = \cos\alpha = 1$, then for a given DDR the diffuseness becomes minimum and equal to
\begin{align}
	\sub{\psi}{pd,w}(\Gamma, 0) = 1-\frac{|\Gamma  + (K/4\pi)|}{\Gamma + 1/Q} = \frac{\sub{\psi}{df,w}}{Q\Gamma + 1}.
\end{align}
A final observation for the axisymmetric case is that contrary to the plane wave case, the opposite of the intensity vector of (\ref{eq:intensepdw}) does not point exactly at the DOA of the plane wave $\Omega_l$ but instead is biased towards the orientation of the beam $\Omega_0$, with the bias being dependent on the power of the diffuse component. From (\ref{eq:intensepdw}) and (\ref{eq:deltapdw}) and the law of sines, the bias $\beta$ towards $\Omega_0$ can be found as
\begin{align}
	\beta(\Gamma, \alpha) = \arcsin \left( \frac{ K \sin\alpha}{4\pi \Delta(\Gamma, \alpha)} \right).
\end{align}

\section{Conclusions}

In this work we present a formulation for estimation of the acoustic intensity, energy density and sound field diffusseness in an angular region determined by a spatial filter. By appropriate definition of the beam pattern, the effect of sources or diffuse noise outside the pattern's main lobe are significantly attenuated and thus contribute less to the energetic estimates. Conversely, sources close to the beam pattern's orientation contribute mostly to the estimates. That allows an energetic analysis for DOA estimation of sound field characterization focused in specific directions and regions. The method is formulated based in the spherical harmonic domain, with respect to the maximum available order of spherical harmonic signals and the coefficients of the spatial filter. Furthermore, an analytical solution is given for the coefficients of the beam patterns that capture the particle velocity vector components as a linear combination of the known spatial filter's coefficients, which constitute a design choice, and sparse deterministic matrices of a special structure which can be precomputed offline. From the directionally-weighted pressure and velocity, the local intensity, energy density and diffuseness can be computed. Finally, the behavior of these quantities is analyzed for reference field conditions with the following conclusions: a) the DOA of a single plane wave field is not altered by the presence of the spatial filter, b) the acoustic intensity does not vanish in the presence of a uniform diffuse field, instead it points to the spatial average of the beam-pattern or to the beam's orientation if
it is axisymmetric, c) the DOA of a single plane wave in the presence of a diffuse field is biased towards the beam's orientation with the bias being dependent on the power of the diffuse field. All the above conclusions can be expressed as a function of the spatial filter's coefficients, and hence they can be used to estimate how much a sound field deviates from these reference conditions.

\bibliographystyle{IEEEbib}
\bibliography{DirectionallyConstrainedIntensity_Politis_Pulkki_2016}

\section*{Appendix: Spherical expansion of a product of spherical functions}
\label{sec:appendix}

Two band-limited spherical functions $f(\Omega)$ of order $N_f$ and $g(\Omega)$ of order $N_g$ have SH coefficients $f_{q'}$ and $g_{q''}$ respectively, with $q'=0,1,...,Q_f-1$ and $q''=0,1,...,Q_g-1$, where $Q_f = (N_f+1)^2$ and $Q_g = (N_g+1)^2$. The spherical function $d(\Omega) = f(\Omega)g(\Omega)$ is the product of the two and of order $N_d=N_f+N_g$, and its SHT is
\begin{equation}
	d_q = \int_{\Omega} \! d(\Omega) Y_q^*(\Omega)\, \mathrm{d}\Omega = \int_{\Omega} \! f(\Omega)g(\Omega) Y_q^*(\Omega)\, \mathrm{d}\Omega.
\end{equation}
By expanding the member functions, we get
\begin{eqnarray}
	\label{eq:gauntproduct}
	d_q &=& \int_{\Omega} \! \left( \displaystyle\sum_{q'=0}^{Q_f-1} f_{q'}Y_{q'}(\Omega) \right) \left( \displaystyle\sum_{q''=0}^{Q_g-1} g_{q''}Y_{q''}(\Omega) \right) Y_q^*(\Omega)\, \mathrm{d}\Omega \nonumber\\
	&=& \displaystyle\sum_{q'=0}^{Q_f-1} \displaystyle\sum_{q''=0}^{Q_g-1} f_{q'} g_{q''} \int_{\Omega} \! Y_{q'}(\Omega) Y_{q''}(\Omega) Y_q^*(\Omega)\, \mathrm{d}\Omega \nonumber\\
	&=& \displaystyle\sum_{q'=0}^{Q_f-1} \displaystyle\sum_{q''=0}^{Q_g-1} G_{q'q''}^q f_{q'} g_{q''}
\end{eqnarray}
where $q = 0,1,...,Q_d-1$ and $G_{q'q''}^q$ are the Gaunt coefficients relating the SH coefficients of the original functions to the coefficients of the product function. The Gaunt coefficients are given by the product of three spherical harmonics $G_{q'q''}^q = \int_{\Omega} \! Y_{q'}(\Omega) Y_{q''}(\Omega) Y_q^*(\Omega)\, \mathrm{d}\Omega$ and can be expressed analytically in terms of the Wigner-3j symbols 
\scriptsize $\left(\begin{array}{ccc}
    n & n' & n'' \\ 
    m & m' & m''
  \end{array}\right)$
\normalsize  as in 
\begin{align}
	G_{q'q''}^{q} &= (-1)^m \sqrt{\frac{(2n+1)(2n'+1)(2n''+1)}{4\pi}} \cdot \notag\\
	&\cdot \left(  \begin{array}{ccc}
    n & n' & n'' \\ 
    m & m' & m''
  \end{array}\right)\, \left(  \begin{array}{ccc}
    n & n' & n'' \\ 
    0 & 0 & 0
  \end{array}\right),
\end{align}
where the $q$ indices are expanded in their order and degree $n,m$ inside the formula. The Wigner-3j symbols can be computed for example by the Racah Formula \cite{wigner3j}.

For the specific case of a product of a beam pattern $w(\Omega)$ with the components of $\vb{n}(\Omega)$ of (\ref{eq:dipole}), the orthogonal dipoles of $\vb{n}(\Omega)$ are only of order $N_n=1$, with coefficients given in (\ref{eq:dipoleSH}). Then the sum of (\ref{eq:gauntproduct}) becomes
\begin{eqnarray}
	w^\mathrm{x}_q &=& \displaystyle\sum_{q'=0}^{Q_w-1} w_{q'} \left( x_1 G_{q'(1)}^q  + x_3 G_{q'(3)}^q \right) \nonumber\\
	w^\mathrm{y}_q &=& \displaystyle\sum_{q'=0}^{Q_w-1} w_{q'} \left( y_1 G_{q'(1)}^q  + y_3 G_{q'(3)}^q \right) \nonumber\\
	w^\mathrm{z}_q &=& \displaystyle\sum_{q'=0}^{Q_w-1} w_{q'} \left( z_2 G_{q'(2)}^q\right)
\end{eqnarray}
which can formulated in the matrix relations of (\ref{eq:velmat})
\begin{eqnarray}
	\vb{w}^\mathrm{x}_{N_w+1} &=& \vb{A}_\mathrm{x} \cdot \vb{w}_{N_w} \nonumber\\
	\vb{w}^\mathrm{y}_{N_w+1} &=& \vb{A}_\mathrm{y} \cdot \vb{w}_{N_w} \nonumber\\	
	\vb{w}^\mathrm{z}_{N_w+1} &=& \vb{A}_\mathrm{z} \cdot \vb{w}_{N_w}.
\end{eqnarray}

\end{document}